\documentstyle[aas2pp4,epsf]{article}
\lefthead{K.S. Dawson \ea }
\righthead{Preliminary Detection of Arcminute Anisotropy}

\slugcomment{Submitted to {\em Astrophysical Journal}}

\def\kl{{\rm{k}\lambda}}
\def\ea{et al.\ }
\def\ah{^{\rm h}}
\def\am{^{\rm m}}
\def\as{^{\rm s}}
\def\pr{^{\prime}}
\def\2pr{^{\prime \prime}}

\def\greatsim{\mathrel{\raise.3ex\hbox{$>$\kern-.75em\lower1ex\hbox{$\sim$}}}}
\def\lesssim{\mathrel{\raise.3ex\hbox{$<$\kern-.75em\lower1ex\hbox{$\sim$}}}}
\def\BDF1{\rm {PC1643+46}}
\def\BDF2{\rm {VLA1312+32}}
\def\BDF3{\rm {PSS0030+17}}
\def\BDF4{\rm {BF0028+28}}
\def\BDF5{\rm {HDF1236+32}}
\def\BDF6{\rm {BF1821+59}}
\def\BDF7{\rm {BF0658+55}}
\def\BDF8{\rm {BF0017+29}}
\def\BDF9{\rm {BF1250+56}}
\def\BDF10{\rm {BF1812+58}}
\def\BDF11{\rm {BF0658+54}}
\begin{document}

\title{A Preliminary Detection of Arcminute Scale Cosmic Microwave Background Anisotropy with 
the BIMA Array}

\author{
K.S.~Dawson\altaffilmark{1},
W.L.~Holzapfel\altaffilmark{1},
J.E.~Carlstrom\altaffilmark{2},\\
M.~Joy\altaffilmark{3},
S.J.~LaRoque\altaffilmark{2}
and E.D.~Reese\altaffilmark{2}}


\altaffiltext{1}{Department of Physics, University of California,
Berkeley CA 94720}
\altaffiltext{2}{Department of Astronomy and Astrophysics, University of Chicago, Chicago
IL 60637}
\altaffiltext{3}{Space Science Laboratory, SD50, NASA Marshall Space Flight Center,
Huntsville AL 35812}
\authoremail{swlh@cfpa.berkeley.edu}

\begin{abstract}
We have used the Berkeley-Illinois-Maryland-Association (BIMA) array
outfitted with sensitive cm-wave receivers to expand  our search for
arcminute scale anisotropy of the Cosmic Microwave Background (CMB).
The interferometer was placed in a compact
configuration to obtain high brightness sensitivity on arcminute scales
over its $6.6\pr$ FWHM field of view.
The sensitivity of this experiment to flat band power peaks at
a multipole of $\ell = 5530$ which corresponds to an angular scale of
$\sim 2^{\pr}$. We present the analysis of a total of $470$ hours of on-source
integration time on eleven independent fields
which were selected based on their 
low IR dust contrast and lack of bright radio sources.
Applying a Bayesian analysis to the visibility data, we find
CMB anisotropy flat-band power $Q_{flat}=6.1^{+2.8}_{-4.8}\,\mu$K at $68\%$ confidence. 
The confidence of a non-zero signal is $76\%$ and we find an upper limit
of $Q_{flat}<12.4\,\mu$K at $95\%$ confidence.
We have supplemented our BIMA observations with concurrent
observations at 4.8 GHz with the VLA to search for and remove point sources.
We find that point sources make an insignificant contribution to the
observed anisotropy.
\end{abstract}


\section{Introduction} \label{sec:intro}

The Cosmic Microwave Background (CMB) radiation carries a wealth of information 
about the early universe. 
In the standard inflationary model, the distribution of matter at the epoch of recombination
leads to small temperature anisotropy of the CMB. Measurement of anisotropy on the largest angular scales
reveals the primordial distribution of matter (\cite{cobe}). Structures which came into
the horizon and were able to collapse 
near the
epoch of recombination lead to anisotropy at degree angular scales. The resulting
CMB anisotropy angular power spectrum at these scales is highly sensitive to the parameters of
the cosmological model. Measurements of degree scale anisotropy  
have been made, for example, to constrain the curvature of the universe
(\cite{miller,debernardis,hanany}). The anisotropy on smaller
angular scales, less than a few arcminutes, will be exponentially damped
to vanishingly small levels due to photon diffusion and from the finite thickness of
the `surface' of last scattering (\cite{Hu97}).
At arcminute scales, so-called secondary anisotropy generated by the reionization of the universe
and the Sunyaev-Zel'dovich effect from galaxy clusters  
should dominate the primary signal (for a review see \cite{Haiman}). 

In this paper we report results from our ongoing program 
using the  Berkeley Illinois Maryland Association (BIMA) interferometer 
to search for arcminute-scale 
CMB anisotropy.
Discussion of the instrument, data reduction, Bayesian maximum likelihood analysis,
expected signals (from both primary and secondary anisotropies) and 
comparison with previous experiments is included in the release of our earlier
results (\cite{Ho00}). 
The field selection and observations are reviewed in \S\ref{sec:obs}.
The results of the Bayesian analysis are presented 
in \S\ref{sec:results} including a discussion of the effects of point source
subtraction. 
Finally, in \S\ref{sec:con}, we summarize the results of the survey.

\section{Observations} \label{sec:obs}
We have used the BIMA array at 28.5~GHz during the summers of 1997, 1998 and 2000 to 
search for CMB anisotropy.
We observed seven independent fields over the course of the first two summers.
In two of these fields we found apparent detections of excess power.
We extended our observations of those two fields in the summer of 2000 and also
added four new fields to the survey.

\subsection{Field Selection} \label{sec:fields}
For observations in the summer of 2000, we selected four new fields, each located
within $2^{\circ}$ of
one of the fields selected in 1998.
We employed such a strategy to make efficient use of observing time allowing
a six hour separation between fields as in the 1998 survey, and to search for
systematic errors which could have resulted in false detections in our previous 
observations.
The new fields were chosen to lie in regions of low dust emission and contrast as 
determined from examination of IRAS $100\,\mu$m maps.
The VLA NVSS~(\cite{NVSS}) and FIRST~(\cite{FIRST}) surveys were then used to 
select regions free of bright point sources 
at $1.4\,$GHz.
In addition, we used the SkyView
digitized sky survey and ROSAT WFC maps
to check for bright optical or x-ray emission which could complicate 
follow-up observations.
The pointing centers for each of the six fields observed in the summer of
2000 are given in 
Table 1.

\subsection{BIMA Observations}
All anisotropy observations were made using the BIMA array at Hat Creek.
The array consisists of nine $6.1$ meter telescopes operating at $28.5\,$GHz,
each with a $6.6\pr$ FWHM primary beam.
In order to track the system gains, each $25$ minute source observation was 
bracketed by a $5$ minute observation of a 
calibrator.
The fluxes of the calibration sources are all referenced to the flux 
of Mars which is uncertain by approximately $4\%$ at $90\%$ confidence 
(see discussion in \cite{Grego}).
Of the total time spent observing, $\sim 60\%$ was spent on source.
The cumulative integration times for each of the 6 fields observed in 2000
are listed in Table 1.
Combined with the observations described in \cite{Ho00}, a total
of 470 hours of on-source integration have been dedicated to this project.

\subsection{VLA Observations and Point Source Results}

\label{sec:VLA}

The compact configuration used for the 2000 Summer BIMA anisotropy observations 
produced high brightness sensitivity but lacked
the spatial dynamic range to distinguish point sources from CMB fluctuations.
To constrain the contribution from point sources to our anisotropy 
measurements, we therefore used the Very Large Array (VLA) to survey
each of the new fields as well as two of the 1998 fields which lacked point source
observations. The $4.8\,$GHz VLA observations were obtained 
within a month of the 2000 Summer BIMA observations.
With an hour and a half of observing time per field, we reached an RMS
flux of
$\sim 25-30\,\mu$Jy at the center of a $9\pr$ FWHM region centered on each of the blank
fields observed with BIMA. 

In the $6$ fields examined with the VLA, we found $18$ point sources with fluxes
adjusted for the attenuation of the primary beam
ranging from $157\,\mu$Jy to $2000\,\mu$Jy.
The average flux of these point sources was $743\,\mu$Jy.

The point source model at $28.5$ GHz is extrapolated from the lower frequency 
VLA data by assuming a spectral index of $\alpha=-0.71$ where
$S_\nu \propto \nu^\alpha$ (\cite{Co98}).
After accounting for attenuation due to the BIMA primary beam,
all of the point sources detected with the VLA are
expected to be near or below the measured RMS flux density achieved
in the BIMA blank field data. 


\section{Results} \label{sec:results}
We have produced and analyzed images for each of the observed fields.
The statistics of the images produced with 
only the short baselines used in the likelihood analysis
are listed in Table 2.
The observed RMS values are 
comparable to those expected from the noise properties of the visibilities.
We also express our results in terms of the RMS Rayleigh-Jeans 
(RJ) temperature fluctuations.

\subsection{Anisotropy Analysis}

We use the method described in \markcite{Ho00} Holzapfel \ea (2000)
to determine the relative likelihoods that the observed fields are described 
by a model for the CMB fluctuations with flat band power $Q_{flat}$. 
We present the results of the data analysis
of the BIMA data both with and without the subtraction of the point sources
extrapolated from the VLA observations.
In Table 3, we show the most likely $Q_{flat}$ for each of the fields observed 
in the summer of 2000 with no point source subtraction.

\begin{figure*}[htb]
\begin{center}
\epsfxsize 4 in
\centerline{\epsfbox{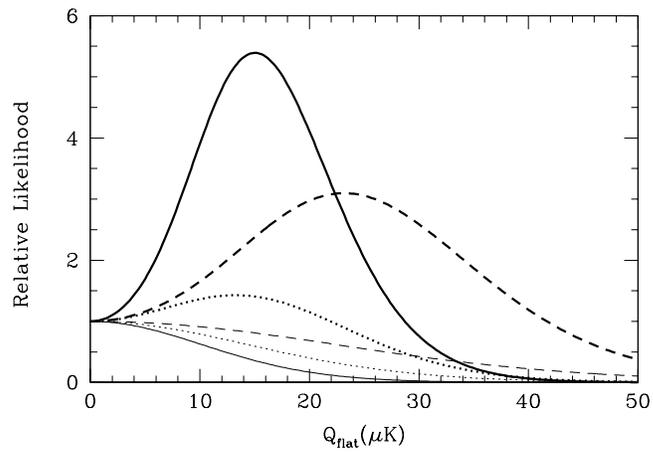}}
\caption{The relative likelihood that the observed signal in each field
is described by flat band power with amplitude $Q_{flat}$ ignoring possible point sources.
The light solid line corresponds to field BDF08, the light dotted line to BDF09,
the light dashed
line to BDF10, the heavy solid line to BDF06, the heavy dotted line to BDF07, and the
heavy dashed line to BDF11.
}
\label{fig:indiv}
\end{center}
\end{figure*}

Figure~\ref{fig:indiv} shows the relative likelihoods as a function
of assumed $Q_{flat}$ in each of the fields 
observed in the summer of 2000 with no point source subtraction.
The results are normalized to unity likelihood for the case of no 
anisotropy signal.
Note that the results for fields BDF6 and BDF7 as displayed in both
Table 3 and Figure~\ref{fig:indiv} also include the data collected
in the summer of 1998.

\begin{figure*}[htb]
\begin{center}
\epsfxsize 4 in
\centerline{\epsfbox{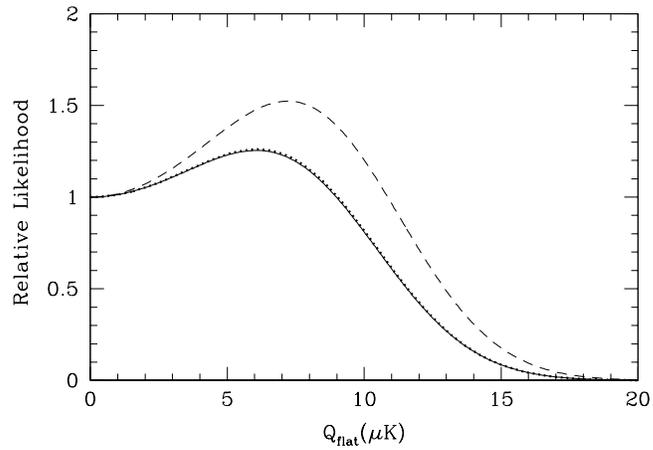}}
\caption{The relative likelihood that the observed signal in the combined fields
is described by flat band power with amplitude $Q_{flat}$.
The solid line corresponds to an analysis ignoring the measured point sources, the
the dotted line is the result of subtracting point sources assuming a spectral index
of -0.71,
and the dashed line is the result of subtracting
measured point sources assuming a flat spectrum.}
\label{fig:all}
\end{center}
\end{figure*}

We estimate the point source contribution to the BIMA results by extrapolating 
the flux of point sources detected with the VLA at $4.8$ GHz to the $28.5$ GHz 
BIMA observation frequency assuming spectral index $\alpha$.
These sources are then removed from the raw data by taking the Fourier transform 
of the point source model modulated by the primary beam response and subtracting 
it directly from the visibility data.
In Figure~\ref{fig:all}, we plot the combined
likelihood for all of the fields under three assumptions for subtracted point 
source model; no point sources, extrapolated fluxes assuming $\alpha=-0.71$, and 
a flat spectrum ($\alpha = 0$) for all sources.
In Table~\ref{tab:qflat}, we list the $68\%$ and $95\%$ confidence intervals 
in $Q_{flat}$ for each of the three point source extrapolations we have considered.
The results are identical before and after the subtraction of the source fluxes 
estimated assuming $\alpha=-0.71$. 
Assuming a flat spectrum, the most likely $Q_{flat}$ slightly increases indicating
that we have overestimated the flux of the point sources and by subtracting 
these sources we are adding excess power to the BIMA data.
It is clear that point sources cannot account for the detected excess power.

Again, as found in \markcite{Ho00} Holzapfel \ea (2000), the joint likelihood for all 
the data, shown in Figure~\ref{fig:all}, peaks at $Q_{flat}>0$.
While the most likely $Q_{flat}$ remains essentially unchanged from the earlier results, 
the addition of the new observations has increased the significance of the detection.
When no point sources are subtracted, the confidence of a non-zero $Q_{flat}$ for the
joint likelihood is $76\%$ which can be compared with $44\%$ for the total of
all data collected prior to the summer 2000 observations.

\subsection{Systematics Check}

Much of the motivation for the new BIMA observations described here was to 
perform tests for systematic errors that could have been responsible for the
previously reported detections. 
We first investigated if the observed excess power could somehow depend
on the position of the source on the sky. 
As described in Section~\ref{sec:fields}, we selected the four new fields to lie within two degrees
of the four fields selected for observation in 1998.
The results for the two 18 hour right ascension fields, BDF6 and BDF10, were compared.  
Observations in 2000 repeated the detection of excess power in BDF6, but none was
found in BDF10 
indicating that the observed excess power in BDF6 is not due to a 
systematic associated with sky position. 
For the two 7 hour right ascension fields, BDF7 and BDF11, the results are not as conclusive;
excess power was detected in both fields. To check for possibility of coherent
power between the fields, e.g., possibly from a local source of interference, 
we analyzed the data from the two fields as if they were from the same
position on the sky. No coherent signal was detected; the level of
the detected power decreased. 
 
We have performed several additional tests of the data to search for 
systematic errors in the fields in which we find excess power.
Since the data on each field was collected over an extended period of time,
we divided the data into blocks of several days and analyzed each block 
independently.
If we were experiencing problems 
over a short period of time, we might expect to find a significantly
larger detection in one of the blocks of data. None was found.
We next checked for the presence of systematic errors as a function of time of day.
For each field, we divided the data into three blocks 
spanning equal intervals in hour angle.
Again,
there was no evidence for anomalous excess power in any of the blocks of data.
Although we see no convincing evidence that our determination of excess power 
is the result of a systematic error, higher sensitivity observations will
be necessary to eliminate this possibility. 
 
 
\section{Conclusion} \label{sec:con}
Over the course of three summers, we have used the BIMA array in a compact 
configuration at $28.5\,$GHz to search for CMB anisotropy in eleven 
independent $6^\prime .6$ FWHM fields.
With these observations, we have made a preliminary detection of arcminute scale
CMB anisotropy.
In the context of an assumed flat band power model for the CMB power spectrum, 
we find $Q_{flat}=6.1^{+2.8}_{-4.8}\,\mu$K at $68.3\%$ confidence 
with sensitivity on scales that 
correspond to an average harmonic multipole $\ell_{eff} = 5530$.
The confidence of a non-zero signal is $76\%$ and we find an upper limit
of $Q_{flat}<12.4\,\mu$K at $95\%$ confidence.
The 28.5~GHz fluxes of the point sources located with the VLA are near or below
the noise level in the 
BIMA images and make no significant contribution when included in the likelihood 
analysis.
A recent search for CMB anisotropy on similar angular scales with the ATCA 
at 8.7~GHz (\cite{Sub00}) was able to place an upper limit of $Q_{flat} < 25\,\mu$K at $95\%$ 
confidence, but this work was believed to be confusion limited by unresolved 
point sources.
We can use our VLA $5\sigma$ flux limit of $\sim 150 \mu\,$Jy at $4.8$ GHz to 
estimate the contribution of unresolved point sources to the excess power we report.
Assuming a universal spectral index of $\alpha=-0.71$, we expect a contribution of 
$Q_{flat} \sim 1.1\,\mu$K due to unresolved point sources in the BIMA data at 28.5~GHz.
Assuming a flat spectral index of $\alpha=0$, we expect a contribution of 
$Q_{flat} \sim 3\,\mu$K as a conservative upper limit.

This work is perhaps the first detection of secondary CMB anisotropy in a region of the
sky not selected for the presence of a known galaxy cluster.
A detection of excess power is not surprising when one considers the level 
of anistropy expected from the SZ effect of  distant clusters of galaxies.
Recent hydrodynamic simulations predict a $Q_{flat}\sim 9.7\,\mu$K (\cite{Vo00}).
Although this value is larger than our most likely fit, it falls well 
within our $68\%$ confidence interval.
Considering the way in which our blank fields are selected, it is not 
surprising to find a lower value for $Q_{flat}$ than is predicted for randomly
selected regions of the sky.
The density of extragalactic radio sources is a tracer 
of large scale structure.
By selecting fields with little point source contamination, we introduce a bias
against finding clusters of galaxies.
If the excess power we have detected is indeed due to the SZ effect in distant 
clusters
of galaxies, deeper observations will resolve the individual 
clusters enabling a new  and powerful probe of large scale structure.


\vskip 20 pt

This paper is dedicated to the memory of Mark Warnock, an electronics
engineer at the BIMA observatory. His expertise and guidance helped
ensure the success of the cm-wave imaging program at BIMA.
We thank the entire staff of the BIMA observatory for their
many contributions to this project. We also thank
Rick Forster, Laura Grego, Daisuke Nagai and Dick Plambeck for assistance with the 
instrumentation and observations. 
This work is supported in part by NASA LTSA grant number NAG5-7986.
The BIMA millimeter array is supported by NSF grant AST 96-13998. We are grateful
for the scheduling of Target of Opportunity time at the VLA in support of this project.

\newpage
\markright{REFERENCES}

\newpage
\markright{TABLES}

\begin{table*}[htb]
\begin{center}
\begin{tabular}{lcccc}
\multicolumn{5}{c}{TABLE 1}\\
\multicolumn{5}{c}{Field Positions and Observation Times}\\\hline\hline
\multicolumn{1}{c}{Fields} & $\alpha\,$(J2000) & $\delta\,$(J2000)  & Observation year(s) & Time (Hrs) \\ \hline
BDF6 & $18\ah\,21\am\,00.0\as$ & $+59^{\circ}\,15\pr\,00\2pr$ & 1998, 2000 &  $81.2$\\
BDF7 & $06\ah\,58\am\,45.0\as$ & $+55^{\circ}\,17\pr\,00\2pr$ & 1998, 2000 & $68.2$\\
BDF8 & $00\ah\,17\am\,30.0\as$ & $+29^{\circ}\,00\pr\,00\2pr$ & 2000 & $34.6$\\
BDF9 & $12\ah\,50\am\,15.0\as$ & $+56^{\circ}\,52\pr\,30\2pr$ & 2000 &  $24.5$\\
BDF10 & $18\ah\,12\am\,37.21\as$ & $+58^{\circ}\,32\pr\,00\2pr$ & 2000 & $14.3$\\
BDF11 & $06\ah\,58\am\,00.0\as$ & $+54^{\circ}\,24\pr\,00\2pr$ & 2000 & $22.1$\\
\end{tabular}
\end{center}
\label{tab:coord}
\end{table*}

\begin{table*}[htb]
\begin{center}
\begin{tabular}{lccccc}
\multicolumn{6}{c}{TABLE 2}\\
\multicolumn{6}{c}{Image Statistics for $u$-$v$ Range $0.63-1.2\,\kl$}\\\hline\hline
\multicolumn{1}{c}{Field} &  Beamsize($^{\2pr}$) & \multicolumn{2}{c}{RMS ($\mu$Jy$\,$beam$^{-1}$)} & \multicolumn{2}{c}{RMS ($\mu$K)} \\
\multicolumn{2}{c}{} & estimated & measured & estimated & measured \\ \hline
BDF6 &  $106.7 \times 118.9$ & $113$ & $166$ & $13.4$ & $19.6$ \\
BDF7 &  $108.0 \times 120.3$ & $130$ & $166$ & $15.0$ & $19.2$ \\
BDF8 & $102.6 \times 116.1$ & $166$ & $133$ & $20.9$ & $16.7$ \\
BDF9 & $101.6 \times 118.9$ & $209$ & $196$ & $26.0$ & $24.3$ \\
BDF10 & $105.3 \times 115.5$ & $275$ & $276$ & $33.9$ & $34.0$ \\
BDF11 & $104.4 \times 115.1$ & $208$ & $279$ & $26.0$ & $34.8$ \\
\end{tabular}
\end{center}
\label{tab:sbimage}
\end{table*}

\begin{table*}[htb]
\begin{center}
\begin{tabular}{lccc}
\multicolumn{4}{c}{TABLE 3}\\
\multicolumn{4}{c}{Most Likely $Q_{flat}$ and Confidence Intervals}\\\hline\hline
$$ & \multicolumn{3}{c}{$Q_{flat}\,(\mu$K)}  \\
Field & Most Likely & $68\%$ & $95\%$ \\\hline
BDF6 & $15.0$ & $\phn 8.2-22.4$ & $0.8-28.6$ \\
BDF7 & $13.2$ & $\phn 3.2-21.2$ & $0.0-31.8$ \\
BDF8 & $\phn 0.0$ & $\phn 0.0-10.4$ & $0.0-21.2$ \\
BDF9 & $\phn 0.0$ & $\phn 0.0-15.0$ & $0.0-30.6$ \\
BDF10 & $\phn 0.0$ & $\phn 0.0-23.2$ & $0.0-47.6$ \\
BDF11 & $23.2$ & $10.8-35.6$ & $0.0-46.2$ \\\hline
\end{tabular}
\label{tab:nopntsub}
\end{center}
\end{table*}

\begin{table*}[htb]
\begin{center}
\begin{tabular}{lcccc}
\multicolumn{5}{c}{TABLE 4}\\
\multicolumn{5}{c}{Analysis of Combined Fields Including Confidence of $Q_{flat}>0$}\\\hline\hline
\multicolumn{1}{l}{} & \multicolumn{3}{c}{$Q_{flat}(\mu{\rm K})$} & Confidence\\
\multicolumn{1}{l}{Point source model} & {Most likely} & $68\%$ & $95\%$ & $Q_{flat}>0$ \\ \hline
none & $6.1$ & $\phn 1.3-8.9$ & $0.0-12.4$ & $76\%$ \\
$\alpha = -0.71$ & $6.1$ & $\phn 1.4-8.9$ & $0.0-12.4$ & $76\%$ \\
$\alpha = 0$ & $7.2$ & $2.4-10.5$ & $0.0-13.2$ & $85\%$ \\
\end{tabular}
\end{center}
\label{tab:qflat}
\end{table*}

\newpage
\markright{figures}

\end{document}